\documentclass[final]{ndes2005}

\usepackage[utf8]{inputenc}
\newcommand{\q}[1]{``#1''}
\newcommand{\citep}[1]{\cite{#1}}
\newcommand{\citet}[1]{\citeauthor{#1} \shortcite{#1}}
\newcommand{\C}{\mathrm{C}}
\newcommand{\R}{\mathrm{R}}
\newcommand{\ph}{\phi}
\newcommand{\ii}{\mathrm{i}}

\begin{document}

%
\catchline{}{}{}{}{}
%

\markboth{C. Allefeld, M. Müller, \& J. Kurths}
{Eigenvalue decomposition as a generalized Synchronization Cluster Analysis}

\title{EIGENVALUE DECOMPOSITION AS A\\
GENERALIZED SYNCHRONIZATION CLUSTER ANALYSIS}

\author{CARSTEN ALLEFELD}

\address{Nonlinear Dynamics Group and Research Unit \q{Conflicting Rules}\\
University of Potsdam, P.O. Box 601553, 14415 Potsdam, Germany\\
\email{allefeld@ling.uni-potsdam.de}
\http{www.agnld.uni-potsdam.de/$\sim$allefeld}}

\author{MARKUS MÜLLER}

\address{Facultad de Ciencias, Universidad Autónoma del Estado de Morelos\\
62210 Cuernavaca, Morelos, Mexico}

\author{JÜRGEN KURTHS}

\address{Nonlinear Dynamics Group and Research Unit \q{Conflicting Rules},
University of Potsdam}

\maketitle

\begin{history}
\received{October 18, 2005}
\revised{January 13, 2006}
\end{history}

\begin{abstract}
Motivated by the recent demonstration of its use as a tool for the detection and characterization of phase-shape correlations in multivariate time series, we show that eigenvalue decomposition can also be applied to a matrix of indices of bivariate phase synchronization strength. The resulting method is able to identify clusters of synchronized oscillators, and to quantify their strength as well as the degree of involvement of an oscillator in a cluster. Since for the case of a single cluster the method gives similar results as our previous approach, it can be seen as a generalized Synchronization Cluster Analysis, extending its field of application to more complex situations. The performance of the method is tested by applying it to simulation data.
\end{abstract}


\section{Introduction}

The eigenvalue decomposition of the equal-time correlation matrix of a set of signals is one of the standard tools of multivariate data analysis (\emph{cf.} \citet{anderson:introduction}). Recently, \citet{mueller:detection} demonstrated the usefulness of the eigenvalue decomposition of the correlation matrix specifically as a tool for the detection of phase-shape correlations in multivariate data sets. They showed that changes in the degree of synchronization in all or a subset of signals are reflected in coordinated changes in the highest and lowest eigenvalues, and that information about the channels involved and the type of their interaction can be obtained from the corresponding eigenvectors.

While the correlation of time series may indicate the synchronization of the oscillators they are obtained from, the physical concept of synchronization refers specifically to the adjustment of the rhythms of oscillators, i.e. to the relative dynamics of their phases rather than their amplitudes \citep{pikovsky:book}. Moreover, there is a regime in the dynamics of coupled chaotic oscillators in which the phase difference is bounded while the amplitudes remain uncorrelated \citep{rosenblum:phase}, called phase synchronization. In this paper, we show that in order to focus the analysis on synchronization relations, it is possible to replace the matrix of correlation coefficients with a matrix of indices of bivariate phase synchronization strength. Combined with an additional step of sorting signals into groups, eigenvalue decomposition can operate as a Synchronization Cluster Analysis, generalizing the previous approach of \citet{allefeld:approach}.

\section{Eigenvalue Decomposition of the Synchronization Matrix}

The correlation matrix $\C$ of a set of data channels $x_i$, $i = 1 \ldots N$, consists of the correlation coefficients $\C_{ij} \in [-1;1]$ between channels. Its eigenvalues $\lambda_k$ and eigenvectors $\vec{v}_k$ are defined by the equation
\begin{equation}
\C ~ \vec{v}_k = \lambda_k ~ \vec{v}_k,
\end{equation}
which in general has $N$ different solutions, $k = 1 \ldots N$. In the following we assume that the eigenvectors are normalized, $|\vec{v}_k| = 1$, and the solutions have been sorted according to the eigenvalues, $\lambda_1 \leq \lambda_2 \leq \ldots \leq \lambda_N$.

The eigenvectors and -values of $\C$ are real-valued, and the eigenvalues are non-negative. Because a matrix becomes the diagonal matrix of its eigenvalues by being transformed into the basis of its eigenvectors and the trace of a matrix is invariant under such a transform, for every correlation matrix holds $\sum \lambda_k = \mathrm{tr} (\C) = N$. In the uncorrelated case this is trivially fulfilled by $\lambda_k = 1$ for all $k$. With a deviation from this, each increase of an eigenvalue above $1$ has to be compensated for by at least one other eigenvalue becoming smaller than $1$, such that this value gives a natural distinction between \q{large} and \q{small} eigenvalues.

The quantification of phase synchronization is based on the instantaneous phase $\ph_i$ of each oscillator $i = 1 \ldots N$. How these phases are determined in the special case is not important here; if the given data are time series, the standard approach is the Hilbert transform for narrowband data, or the Morlet wavelet transform for broadband signals (for a discussion, see \citet{pikovsky:phase} or \citet{allefeld:phd}, Sec.~3.2). The statistical strength of phase synchronization of two oscillators $i$ and $j$ can then be defined as the \q{peakedness} of the distribution of the phase difference $\ph_j - \ph_i$; here we use the measure
\begin{equation}
\R_{ij} = \left | {1 \over n} \sum_{l = 1}^n \exp \left ( \ii ~ (\ph_{jl} - \ph_{il}) \right ) \right |,
\end{equation}
where $l = 1 \ldots n$ enumerates the realizations in the given sample. For the continuum from no to perfect phase synchronization, this measure takes on values from $0$ to $1$. $\R$ can be seen as the modulus of the complex correlation coefficient of signals $x_i = \exp(\ii ~ \ph_i)$, and its decomposition shares the properties given above for $\C$.

Since $\R$ is a nonlinear measure, in this case the eigenvalue decomposition can no longer be interpreted with regard to a linear transform of data channels into source channels. But still the result of the decomposition can be used as a means to analyze the structure of synchronization relations. We will demonstrate this with two basic, artificially constructed examples of synchronization matrices.

\begin{figure}
\centerline{\includegraphics[scale=0.8]{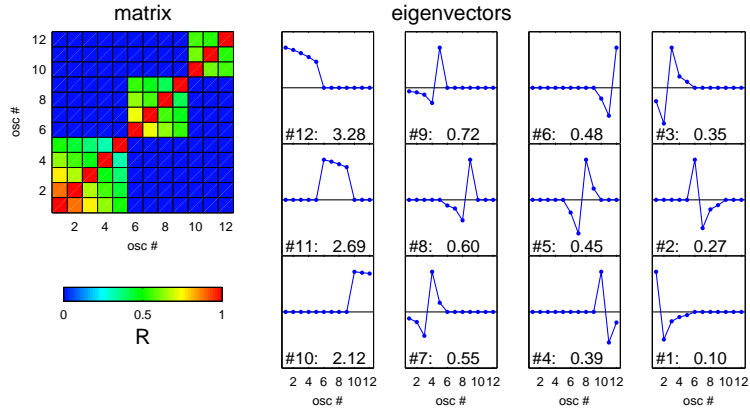}}
\vspace{-1em}
\caption{Left: Synchronization matrix consisting of three clusters of oscillators. Right: Its eigenvectors and -values. Eigenvectors corresponding to eigenvalues $>1$ describe the cluster structure.}
\label{attenuated}
\end{figure}

\begin{figure}
\centerline{\includegraphics[scale=0.8]{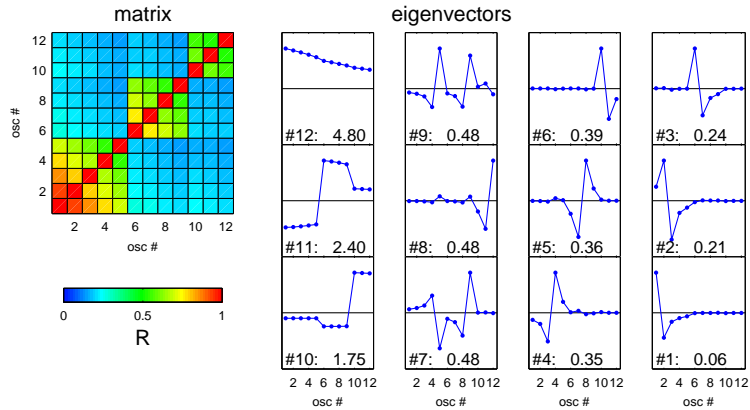}}
\vspace{-1em}
\caption{Synchronization matrix consisting of three clusters with additional inter-cluster synchronization. The cluster structure appears in the eigenvectors for $\lambda_k > 1$, but only in different superpositions.}
\label{synchronized}
\end{figure}

Figure~\ref{attenuated} shows the synchronization matrix for a system consisting of three clusters of synchronized oscillators (with no synchronization between clusters and a different degree of involvement of each oscillator in its cluster) along with the result of the eigenvalue decomposition. There are three eigenvalues larger than one, and the corresponding eigenvectors describe clearly the extent of the clusters as well as the degree of involvement of the individual oscillators. There is also a correspondence between the size and strength of internal synchronization of the clusters and the three eigenvalues. In contrast, the remaining eigenvectors seem not to contribute to the description of the synchronization clusters. The interpretation of the eigenvalue decomposition of $\R$ therefore has these aspects: 1)~Synchronization clusters are identified by eigenvalues $\lambda_k > 1$. The eigenvalues themselves quantify the \emph{strength} of the clusters. 2)~For each cluster, the corresponding eigenvector describes its internal structure. Because $\sum_i v_{ik}^2 = 1$, the index $v_{ik}^2$ quantifies the relative involvement of channel (oscillator) $i$ in cluster $k$. 3)~Combining both, the \q{absolute} involvement of channel $i$ in cluster $k$ can be quantified by the \emph{participation index} $\lambda_k v_{ik}^2$.

Figure~\ref{synchronized} gives the result for a matrix consisting of the same three clusters, but with additional inter-cluster synchronization. Because of the coupling between them, the three clusters no longer appear in separate components of its eigenvalue decomposition. There are still three $\lambda_k > 1$, but the eigenvectors consist of superpositions of the clusters. To account for this, the oscillators belonging to the three clusters have to be identified explicitely.

\begin{figure}
\centerline{\includegraphics[scale=0.8]{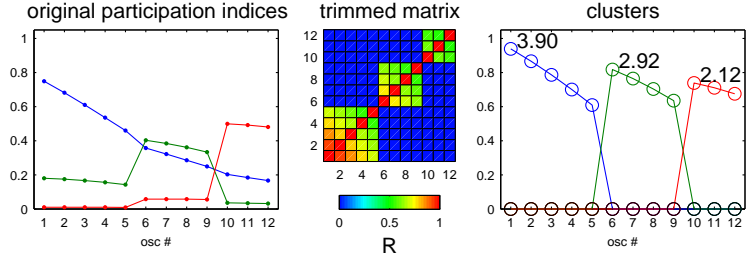}}
\vspace{-1em}
\caption{Synchronization Cluster Analysis. Left: Participation indices corresponding to the $\lambda_k > 1$ for the synchronization matrix of Fig.~\ref{synchronized}. Oscillators are attributed to that cluster for which its participation index is maximal: blue, green, or red. Center: Synchronization matrix with removed inter-cluster synchronization. Right: Result of the eigenvalue decomposition of the trimmed matrix; shown are the participation indices and cluster strengths.}
\label{synchronized:ca}
\end{figure}

This can be done by means of the participation indices; the procedure is illustrated in Fig.~\ref{synchronized:ca}. Each oscillator is attributed to that cluster for which its participation is maximal. In this way, the three clusters consisting of oscillators \#1--5 (blue), 6--9 (green), and 10--12 (red) are correctly identified. In a second step, all of the indices for inter-cluster synchronizations are set to zero, and the eigenvalue decomposition is repeated on the trimmed matrix. For the result of this decomposition, the interpretation given above is valid again.

\section{Generalized Synchronization Cluster Analysis}

The procedure described in the last section is an approach to synchronization cluster analysis. The method identifies clusters of synchronized oscillators and quantifies the strength of the clusters as well as the degree of involvement of each oscillator in its cluster. In a previous paper, \citet{allefeld:approach} introduced another approach that was limited to a single cluster of synchronization; it assumed that all of the oscillators belong to the same cluster, and focused on the quantification of the degree of oscillator participation. The algorithm derived from the observation that under relatively general conditions the synchronization indices within a cluster can be written as the product of factors $\R_{i\C}$,
\begin{equation}
\R_{ij} = \R_{i\C} ~ \R_{j\C}
\quad \textrm{for} \quad i \neq j
\quad (\R_{ii} = 1),
\end{equation}
which can be interpreted as the synchronization strength between oscillator $i$ and the cluster itself, its \q{to-cluster synchronization strength}.

\begin{figure}
\centerline{\includegraphics[scale=0.8]{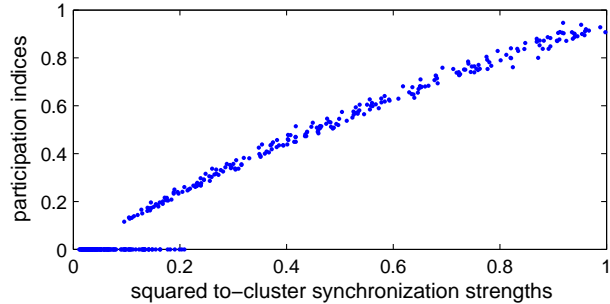}}
\vspace{-1em}
\caption{Relation between the result of the single-cluster analysis (horizontal scale: estimate of $\R_{i\C}^2$) and the par\-ti\-ci\-pa\-tion indices given by eigenvalue decomposition (vertical scale: $\lambda_N v_{iN}^2$) in 20 simulation runs.}
\label{scaCluster}
\end{figure}

We investigated the relationship between the results of the two methods by applying them to simulation data that conforms to the presupposition of a single factorizable synchronization cluster. For $N = 20$ oscillators, the $\R_{i\C}$ were drawn randomly from the uniform distribution over $[0;1]$; then the $\R_{ij}$ were calculated based on $100$ samples of the phase difference from a wrapped normal distribution (\emph{cf.} \citet{allefeld:approach}). Figure~\ref{scaCluster} shows the result for the relation between the output of the earlier method and the participation indices for the strongest cluster. Though the mathematical background of both methods is clearly different, the plot shows an almost functional dependency, which can be roughly described by $\R_{i\C}^2 = \lambda_N v_{iN}^2$. The region of zero participation indices for $\R_{i\C}^2$ below about $0.2$ comes from the attribution of weakly synchronized oscillators to another cluster by the new method; in this range the observed value can no longer be reliably distinguished from no synchronization for the given sample size. Since there is a clear relationship between the results of both methods in the case where the assumptions of the earlier method hold, the cluster analysis based on the eigenvalue decomposition of the synchronization matrix can be seen as a generalization of our previous approach to synchronization cluster analysis.

\section{Application to Simulated Phase Synchronization}

To check the performance of the new method, we applied it to data obtained from the numerical simulation of a system that is known to exhibit clusters of phase synchronization, previously investigated by \citet{osipov:phase} (see Eq.~1 \& Sec.~IV~A). The system consists of a chain of $20$ phase-coherent Rössler oscillators with a diffusive coupling in the $y$-component of strength $\epsilon = 0.007$. The natural frequencies of the oscillators increase linearly along the chain, in the range $\omega = 1 \ldots 1.004$. Phases were defined as the angle in the $x/y$-plane. To also test for the influence of small sample size and noise, we additionally applied the algorithm to the synchronization matrix obtained from data reduced to $100$ approximately independent samples with $50\%$ white noise added to the time series data.

\begin{figure}
\centerline{\includegraphics[scale=0.8]{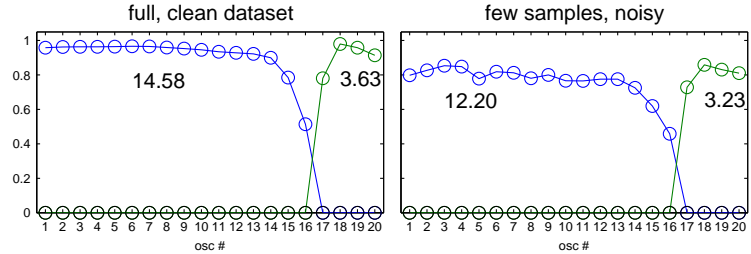}}
\vspace{-1em}
\caption{Analysis results (participation indices and cluster strengths) for a chain of nonidentical Rössler oscillators. Left: Based on original simulation data. Right: Data reduced to $100$ samples, plus $50\%$ measurement noise.}
\label{rchain}
\end{figure}

The result is shown in Fig.~\ref{rchain}. The analysis identifies two clusters of synchronized oscillators (\#1--16 \& 17--20). There is an almost constant high participation of oscillators in their cluster, resulting in cluster strengths close to the number of involved oscillators. Decreased participation occurs near the border between the clusters, where the coupling along the chain pulls oscillators away from the common dynamics of their group. The result for reduced sample size and measurement noise is very similar to the original. Though participation indices and cluster strengths are slightly smaller due to the noise, the same two clusters are clearly identified. This result indicates that the cluster analysis based on eigenvalue decomposition is relatively robust against small sample size and noise.

\section*{Conclusion and Acknowledgements}

We introduced a new approach to synchronization cluster analysis based on eigenvalue decomposition. The new method can be seen as a generalization of our previous approach, extending its field of application to situations including multiple clusters. The algorithm was tested on simulation data and shown to be robust against small sample size and noise. Future work will be on the use of the method to investigate synchronization patterns in EEG data. A \textsc{Matlab} implementation of the algorithm can be obtained from the corresponding author.---This work has been supported by Deutsche Forschungsgemeinschaft.

\end{document}